\documentclass[twocolumn]{aastex631}

\begin{document}

\title{Follow-up observations of apparently one-off sources from the Parkes telescope}

\email{sbzhang@pmo.ac.cn, yangxuan@pmo.ac.cn}

\author{Songbo Zhang}
\affiliation{Purple Mountain Observatory, Chinese Academy of Sciences, Nanjing 210023, China}
\affiliation{CSIRO Space and Astronomy, Australia Telescope National Facility, PO Box 76, Epping, NSW 1710, Australia}

\author{Xuan Yang}
\affiliation{Purple Mountain Observatory, Chinese Academy of Sciences, Nanjing 210023, China}
\affiliation{School of Astronomy and Space Sciences, University of Science and Technology of China, Hefei 230026, China}
\affiliation{CSIRO Space and Astronomy, Australia Telescope National Facility, PO Box 76, Epping, NSW 1710, Australia}



\begin{abstract} 
A small fraction of fast radio bursts (FRBs) have been observed with multiple bursts, whereas most Galactic sources emitting radio pulses are known to repeat. Here we present the results of follow-up observations of two FRBs and four rotating radio transients (RRATs). Among these, only one RRAT has been observed with repeating pulses, with an estimated period of around 1.297047\,s. For comparison, we reanalysed the Parkes archival follow-up observations in CSIRO's data archive for all apparently one-off sources discovered by the Parkes telescopes, including 13 RRATs and 29 FRBs. In total, 3 RRATs are suggested to be repeaters, but no repeating signals were detected from the other sources. Reporting details of the non-detection observations for the apparently one-off sources would help investigate their origins, and catastrophic scenarios are worth proposing for both extragalactic and Galactic sources. 
\end{abstract}


\section{Introduction} \label{sec:intro}

Before a population of fast radio bursts (FRBs) was reported~\citep{Thornton13}, sources emitting radio pulses were generally thought to be repeating~\citep{McLaughlin06}, and in principle, a transient was normally considered a convincing astrophysical source only when it had been detected in at least two observations~\citep{Lorimer04handbook}.
This trend led to a long debate about the astrophysical origin of the ``Lorimer burst''~\citep{Lorimer07, Burke-Spolaor11, Keane12, Bannister14}, which no repeat pulse was detected in up to 90 hours of follow-up observations~\citep{Lorimer24}. However, \cite{Thornton13} ended this debate, and worldwide FRB-hunting instruments~\citep{Masui15,Spitler16,Shannon18}, especially the Canadian Hydrogen Intensity Mapping Experiment (CHIME) telescope~\citep{CHIME21}, have provided a large FRB sample including over 600 sources. This instantaneously changed the narrative of apparently one-off sources: although it is widely discussed whether all FRBs repeat, the fact is only around 10\% of all currently known FRBs exhibit repeat bursts~\citep{CHIME23}.

Besides the extragalactic source like FRBs, some apparently one-off Galactic sources were also discussed by~\cite{Keane16}: twelve sources labelled as rotating radio transients (RRATs) have never been observed to show repeating pulse, and the previous justification of astrophysical radio source ~\citep{Lorimer04handbook} potentially leads to bias of the reported single pulse Galactic events.
However, it is notable that only the repeating sources could be proven to repeat, whereas one-off sources cannot be definitively classified. Therefore, for any apparently one-off source, it is important to present its detection significance and pulse rate constraint, which is relatively lacking, especially for Galactic sources.

CSIRO's data archive\footnote{CSIRO Data Access Portal, DAP, \url{https://data.csiro.au}}~\citep{Hobbs11} provides the majority of publicly available Parkes high-time resolution datasets and is excellent for analysing the results of follow-up observations of apparently one-off sources discovered by the Parkes telescope.  
The Parkes telescope dominated early FRB discoveries~\citep{Lorimer07, Keane12, Thornton13, Burke-Spolaor14, Petroff15, Ravi15, Champion16, Ravi16, Keane16Nature, Petroff17} and has detected 30 FRB sources to date~\citep{Bhandari18, Bhandari18Atel, Zhang19, Oslowski19, Price19, Petroff19, Zhang20}. However, only one of these FRBs was observed to repeat through the highly sensitive Five-hundred-meter Aperture Spherical radio Telescope (FAST)~\citep{Luo20}.
Meanwhile, 13 RRATs discovered by Parkes were reported to be apparently one-off~\citep{Keane11, Burke-Spolaor10, Burke-Spolaor11, Zhang20}.

In this letter, we present the results of our follow-up observations of the two FRBs and four RRATs we discovered from the Parkes archive~\citep{Zhang19, Zhang20}, as well as the publicly archival observations of other apparently one-off FRBs and RRATs discovered by Parkes. 
In Section~\ref{sec:obs}, we describe the details of the observations and data reduction. The results from these datasets are presented in Section~\ref{sec:results}. We discuss the results and conclude in Section~\ref{sec:con}.

\section{Observation and data reduction} \label{sec:obs}

\begin{table*}
\footnotesize
\caption{{\bf Parkes public archival follow-up observations of apparently one-off RRATs and FRBs discovered by the Parkes telescopes.} The columns are as follows: (1) RRAT names based on the J2000 coordinate and FRB names based on the initial detection date. $^*$ indicates that the file containing the initial detection is publicly accessible on DAP, and the presented properties are derived through our re-processing. (2, 3) J2000 right ascension and declination of the pointing centre for the detected beam. (4) The localization uncertainties in units of arcminutes. (5,6) The S/N and DM of the initial detection. (7, 8, 9) The archival Parkes follow-up observations of Multibeam, UWL and combined two (hereafter referred to as total observation). Notably, several large survey datasets are awaiting transfer and storage in DAP, the current Parkes archive is incomplete. $^+$ indicates the value is the sum of the referred published follow-up observation and the archival observations later than its published time. They are $^{R_1}$~\citep{Burke-Spolaor11HTRU} of 0.77\,hr for J1135$-$49 and 0.14\,hr for J1541$-$42, $^{R_2}$~\citep{Lorimer24} of 90\,hr, $^{R_3}$~\citep{Ravi15} of 78\,hr, $^{R_4}$~\citep{Petroff15} of 14.7\,hr, $^{R_5}$~\citep{Petroff17} of 10\,hr, $^{R_6}$~\citep{Keane16Nature} of 5.84\,hr, $^{R_7}$~\citep{Bhandari18} of 10\,hr for FRB 150610A, 3\,hr for FRB 151206A and 9.2\,hr for FRB 160102A, $^{R_8}$~\citep{Ravi16} of 90\,hr. (10,11,12) Date of the initial detection, the earliest and latest observation available in DAP.}
    \centering
    \begin{tabular}{l|ccccccccccc}
    \hline
    \hline
    Name  & Pointing RA  & Pointing Dec & L$_{\rm unc}$ & S/N  & DM               &          T$_1$ & T$_2$  & T$_{\rm tot}$  & D$_{\rm det}$ & D$_{\rm ear}$  & D$_{\rm lat}$  \\
          & (J2000)       & (J2000)        & ($^{\prime}$)              &      & (pc\,cm$^{-3}$)  & (hr)  & (hr)   &   (hr)         &   (MJD)       & (MJD)          & (MJD)          \\
    \hline
    J0845$-$36$^*$    & 08:45:07.2 & -36:05:31.5 & 14  & 7.0  & 29(2)   & 1.75 & 0  &  1.75 & 52601 & 50993 & 55240  \\
    J0923$-$31        & 09:23:38.1 & -31:57:17.8 & 14  & 9.0  & 72(20)  & 0.89 & 0  &  0.89 & - &  51197 & 55649  \\
    J1135$-$49        & 11:35:56   & -49:25:31   & 14 &$\sim$10 & 114(20)& 0.77$^{R_1}$ & 0 &  0.77$^{R_1}$ & - & - & -  \\
    J1311$-$59$^*$    & 13:11:36.5 & -59:19:12.3 & 14  & 10.6 & 149(4)  & 3.88 & 0  &  3.88 & 51467 & 51467 & 55108  \\
    J1328$-$58$^*$    & 13:28:55.8 & -58:54:05.9 & 14  & 19.4 & 213(2) & 2.09 & 3.69  & 5.78  & 51558 & 51558 & 59195  \\
    J1541$-$42        & 15:41:55   & -42:18:50   & 14 &$\sim$10 & 60(10) & 0.14$^{R_1}$ & 0 &  0.14$^{R_1}$ & - & - & -  \\
    J1605$-$45$^*$    & 16:05:35.7 & -45:45:05.2 & 14  & 11.4 & 65(2)   & 0.83 & 0.99  & 1.82 & 51716 & 51716 & 59100  \\
    J1610$-$17$^*$    & 16:10:24.2 & -17:52:13.0 & 14  & 8.7  & 53(3)   & 1.56 & 0  &  1.56 & 52331 & 52331 & 57233  \\
    J1649$-$46$^*$    & 16:49:47.8 & -46:13:46.4 & 14  & 10.1 & 404(10) & 3.68 & 0  &  3.68 & 51457 & 51457 & 59146  \\
    J1905$-$01$^*$    & 19:05:49.0 & -01:26:42.1 & 14  & 10.2 & 103(3)  & 0.88 & 0.75  & 1.63& 51948 & 51948 & 59199   \\    
    \hline
FRB010125A$^*$& 19:06:53.0& -40:37:14.4& 14 & 17.9 & 786.5(3)  & 0.23 & 0 & 0.23 & 51934 & 51934 & 56846 \\
FRB010305A$^*$& 04:57:19.5& -52:36:24.7& 14 & 10.2 & 350(5)    & 0.07 & 9.87 & 9.94 & 51973 & 51973 & 59197 \\
FRB010312A$^*$& 05:26:54.9& -64:56:19.2& 14 & 11.0 & 1163(20)  & 2.38 & 3.89 & 6.27 & 51980 & 51979 & 59123 \\
FRB010621A$^*$& 18:52:05.1& -08:29:35.0& 14 & 15.8 & 749(10)   & 25.52& 0    &25.52 & 52081 & 52081 & 55659 \\
FRB010724A$^*$& 01:18:06.0& -75:12:18.7& 14 & 32.0 & 373(3)    & 90$^{R_2}$ & 0 & 90$^{R_2}$ & 52114 & 52079 & 57650 \\
FRB090625A$^*$& 03:07:47.2& -29:55:35.9& 14 & 25.2 & 899.6(1)  & 2.64 & 0 & 2.64 & 55007 & 55007 & 57327 \\
    FRB110214A& 01:21:17  & -49:47:11  & 14 & 13   & 168.8(5)  &30.32 &0 & 30.32& 55606 & 57188 & 57934 \\
    FRB110220A& 22:34:38  & -12:23:45  & 14 & 49 & 944.38(5)   & - & 0 & - & 55612 & - & - \\
    FRB110626A& 21:03:43  & -44:44:19  & 14 & 11 & 723(0.3)    & - & 0 & - & 55738 & - & - \\
    FRB110703A& 23:30:51  & -02:52:24  & 14 & 16 & 1103.6(7)   & - & 0 & - & 55745 & - & - \\
    FRB120127A& 23:15:06  & -18:25:38  & 14 & 11 & 553.3(3)    & 0.31 & 0 & 0.31 & 55953 & 57192 & 58125 \\
    FRB121002A& 18:14:47  & -85:11:53  & 14 & 16 & 1629.18(2)  & 46.5 & 0 & 46.5 & 56202 & 52419 & 58560 \\
    FRB130626A& 16:27:06  & -07:27:48  & 14 & 20 & 952.4(1)    & 0.23 & 0 & 0.23 & 56469 & 52457 & 57864 \\
    FRB130628A& 09:03:02  & +03:26:16  & 14 & 29 & 469.88(1)   & 0.16 & 0 & 0.16 & 56471 & 57957 & 57957 \\
    FRB130729A& 13:41:21  & -05:59:43  & 14 & 14 & 861(2)      & 11.17 &0 & 11.17 & 56502 & 57913 & 58335 \\
    FRB131104A& 06:44:10  & -51:16:40  & 14 & 30 & 779(1)      & 85.31$^{+R_3}$ & 0 & 85.31$^{+R_3}$ & 56600 & 52178 & 58330 \\
    FRB140514A& 22:34:06.2& -12:18:46.5& 14 & 16 & 562.7(6)    & 27.64$^{+R_4}$ & 0 & 27.64$^{+R_4}$ & 56791 & 57327 & 58142 \\
    FRB150215A& 18:17:27  & -04:54:15  & 14 & 19 & 1105.6(8)   & 10$^{R_5}$ & 0 & 10$^{R_5}$ & 57068 & 51256 & 55305 \\
    FRB150418A& 07:16:30.9& -19:02:24.4& 14 & 39 & 776.2(5)   & 50.73$^{+R_6}$ & 5.79 & 56.52$^{+R_6}$ & 57130 & 52139 & 59178 \\
FRB150610A$^*$& 10:44:27.0& -40:05:23.2& 14 & 16.9  & 1593.1(4)   & 16.93$^{+R_7}$ & 1.50 & 18.43$^{+R_7}$ & 57183 & 51931 & 58620 \\
    FRB150807A& 22:40:24.4& -53:15:46.8& 1.8& 50  & 266.5(1)    & 90$^{R_8}$ & 0 & 90$^{R_8}$ & 57241 & 54897 & 57867 \\
    FRB151206A& 19:21:25  & -04:07:54  & 14 & 10 & 1909.8(6)   & 11.57$^{+R_7}$ & 0 & 11.57$^{+R_7}$ & 57362 & 51257 & 58560 \\
    FRB151230A& 09:40:50  & -03:27:05  & 14 & 17 & 960.4(5)    & 37.16$^{+R_7}$ & 0 & 37.16$^{+R_7}$ & 57386 & 52256 & 58570 \\
    FRB160102A& 22:38:49  & -30:10:50  & 14 & 16 & 2596.1(3)   &27.00$^{+R_7}$ & 0 & 27.00$^{+R_7}$ & 57389 & 57734 & 58559 \\
    FRB171209A& 15:50:25  & -46:10:20  & 14 & 35.8 & 1457.4(3)  & 3.53 & 0 & 3.53 & 58096 & 51063 & 59105 \\
    FRB180309A& 21:24:43.8& -33:58:44.5& 14 &112.8 & 263.42(1)  & 1.86 & 0 & 1.86 & 58186 & 51935 & 53308 \\
    FRB180311A& 21:31:33.4& -57:44:26.7& 14 &15.3& 1570.9(5)  & 0.83 & 0 & 0.83 & 58188 & 54897 & 57204 \\
    FRB180714A& 17:46:12  & -11:45:47  & 14 &19.8  & 1467.9(3)  & 0.23 & 0 & 0.23 & 58313 & 51068 & 55059 \\
FRB180923B$^*$& 15:10:55.4& -14:06:10.2& 14 &13  & 548(3)     & 4.63&3.59& 8.22 & 58384 & 58383 & 59870 \\
    \hline
    \end{tabular}
    \label{tab:property}
\end{table*}

We carried out follow-up observations using the Ultra-Wideband Low (UWL) receiver~\citep{Hobbs20UWL} of the Parkes telescope for two FRBs: FRB 010305A and FRB 010312A, and four RRATs: SPC 000115 (hereafter J1328$-$58), SPC 000621 (hereafter J1605$-$45), SPC 001122 (hereafter J0808$-$32) and SPC 010208 (hereafter J1905$-$01). 
Our observing pointing followed the pointing centres of the detected beams~\citep{Zhang19, Zhang20} of the Parkes Multibeam receiver system~\citep{Staveley-Smith96MB}. 
The UWL system covers frequencies from 704\,MHz to 4032\,MHz. Data were recorded with 2-bit sampling every 64 $\mu$s, in each of the 0.5\,MHz wide frequency channels.  
Data were coherently de-dispersed at the dispersion measures (DMs) of the initial detections (DM$_{\rm det}$) with only one polarization being recorded.
However, when any source was observed to be repeating, full Stokes information was collected. A 2-minute noise diode signal was also recorded for polarization calibration before each observation.
In total, we observed J1328$-$58 for 3.69\,hr, J1605$-$45 for 0.99\,hr, J0808$-$32 for 15.62\,hr, J1905$-$01 for 0.75\,hr, FRB 010305A for 9.87\,hr and FRB 010312A for 3.89\,hr, respectively\footnote{The observations were conducted as parts of several proposals focused on searching for repeating FRBs and RRATs, as well as during available green time of Parkes. The newly discovered repeating source J0808$-$32 (see Section~\ref{sec:results_repeating}) has been observed the longest. FRBs, which were prioritized in these proposals, and the J1328$-$58, which had the highest initial detected S/N among our targets, were allocated relatively long tracking times. Due to the limited total allocated observation time, the two one-off Galactic sources, J1605$-$45 and J1905$-$01, were only observed from about 1\,hr each.}. 
In addition to our observations, we also obtained archival observations with pointing within 7\,arcmin ($\sim 0.5 \times$ the half power beamwidth of the Multibeam receiver system) of the pointing centres of the detected beams (hereafter referred to as follow-up observations) for 13 RRATs and 29 FRBs discovered by the Parkes telescope and reported as apparently one-off sources. 
Since all these events were detected by the Parkes Multibeam receiver system, we collected only the archival Multibeam datasets and UWL datasets, the latter were supposed to have higher sensitivity and larger field of view (FOV) than the Multibeam datasets~\citep{Hobbs20UWL}.  
Except for three sources that proved to be repeating (see details in Section~\ref{sec:results_repeating}), the details of the follow-up observations for other sources are listed in Table~\ref{tab:property}.

Data collected from the Multibeam and UWL receivers were processed using search pipelines based on the pulsar/FRB single pulse searching package \emph{\sc presto}~\citep{Ransom01}. 
We processed the full band data from the Multibeam receiver following a well-used pipeline~\citep{Zhang20}, but divided the UWL data into a series of sub-bands ranging from 128 to 3328\,MHz based on a tiered strategy~\citep{Kumar21_11a}.  
Strong narrow-band and short-duration broadband radio frequency interference (RFI) were identified and marked using the \emph{\sc presto} routine \emph{\sc rfifind}.
Data were then dedispersed in a range of DM values of DM$_{\rm det}$ $\pm$ 20 pc\,cm$^{-3}$, with a step of $0.1$\,cm$^{-3}$. Single pulse candidates with signal-to-noise ratio (S/N) greater than 7 were recorded and visually inspected.  
To minimize the false positives caused by statistical noise fluctuations or structured RFI, a candidate was identified as a detected pulse only if it exhibited a plausible sweep in the dedispersed frequency-time plane. For multibeam data, the candidate also had to be detected in no more than three adjacent beams.
For the data with full Stokes information, the de-dispersed polarization data were calibrated using the {\sc psrchive} software package~\citep{Hotan04, Straten12} with correction for differential gain and phase between the receivers achieved through the injection of a noise diode signal before each observation. 
Rotation measures (RMs) for all detected pulses were measured using the {\sc rmfit} program in the package {\sc psrchive}~\citep{Hotan04, Straten12}, searching for a peak in the linearly polarized flux $L = \sqrt{Q^2 + U^2}$, within the RM range from $-4000$ to 4000\,rad\,m$^{2}$, with a step of $1$\,rad\,m$^{2}$. {\sc rmfit} corrects for Faraday rotation for each trial RM, producing a total linear polarization profile and an RM spectrum. A Gaussian fit was then applied to determine the optimal RM along with its 1$\sigma$ uncertainty.

\section{Results} \label{sec:results}

\subsection{Repeating sources}\label{sec:results_repeating}

\begin{table*}
\footnotesize
\caption{{\bf Properties of three repeating sources.} The columns are as follows: (1, 2, 3, 4, 5) Names, J2000 right ascension and declination, localization uncertainties, and DM of the reported one-off detections. (6, 7, 8, 9) Names, J2000 right ascension and declination, and DM of the related repeating sources. (10) the differences in localization between the reported one-off detections and the related repeating sources.}
    \centering
    \begin{tabular}{ccccc|cccc|c}
    \hline
    \hline
    Name of& Pointing RA  & Pointing Dec & L$_{\rm unc}$  &      DM         & Name of related    & RA      & Dec     &    DM           & L$_{\rm diff}$ \\
    one-off detection & (J2000) & (J2000) & ($^{\prime}$) & (pc\,cm$^{-3}$) & repeating source   & (J2000) & (J2000) & (pc\,cm$^{-3}$) &  ($^{\prime}$) \\
    \hline
    SPC001122    & 08:08:10.6 & -32:18:11.0 & 14  & 136(10)   & J0808$-$32  &     -       &     -     &127.2(6)&  -  \\
    SPC991113    & 17:39:49.6 & -25:13:16.2 & 14  & 203(26)  & J1739$-$2521 & 17:39:32.83 & -25:21:02 &186.4   & 9.5 \\
    J1709$-$43   & 17:09:47   & -43:54:43   & 14  & 228(20)  & J1709$-$4401 & 17:09:41.39 &-44:01:11.2&225.8(4)& 6.5 \\
    \hline
    \end{tabular}
    \label{table:repeating}
\end{table*}

\begin{figure*}
\begin{center}
\begin{tabular}{ccc}
\includegraphics[width=5.2cm,angle=0]{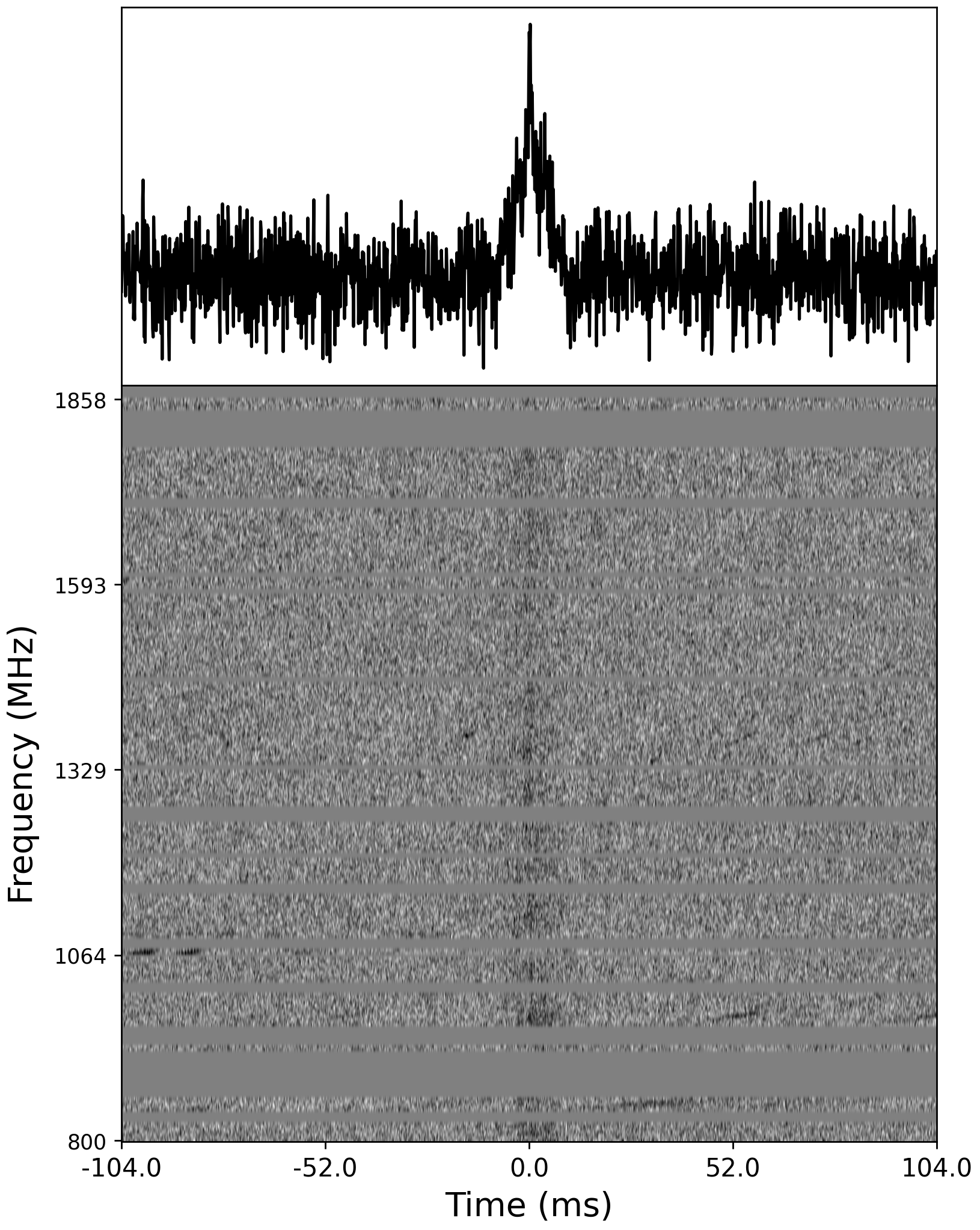} &
\includegraphics[width=5.2cm,angle=0]{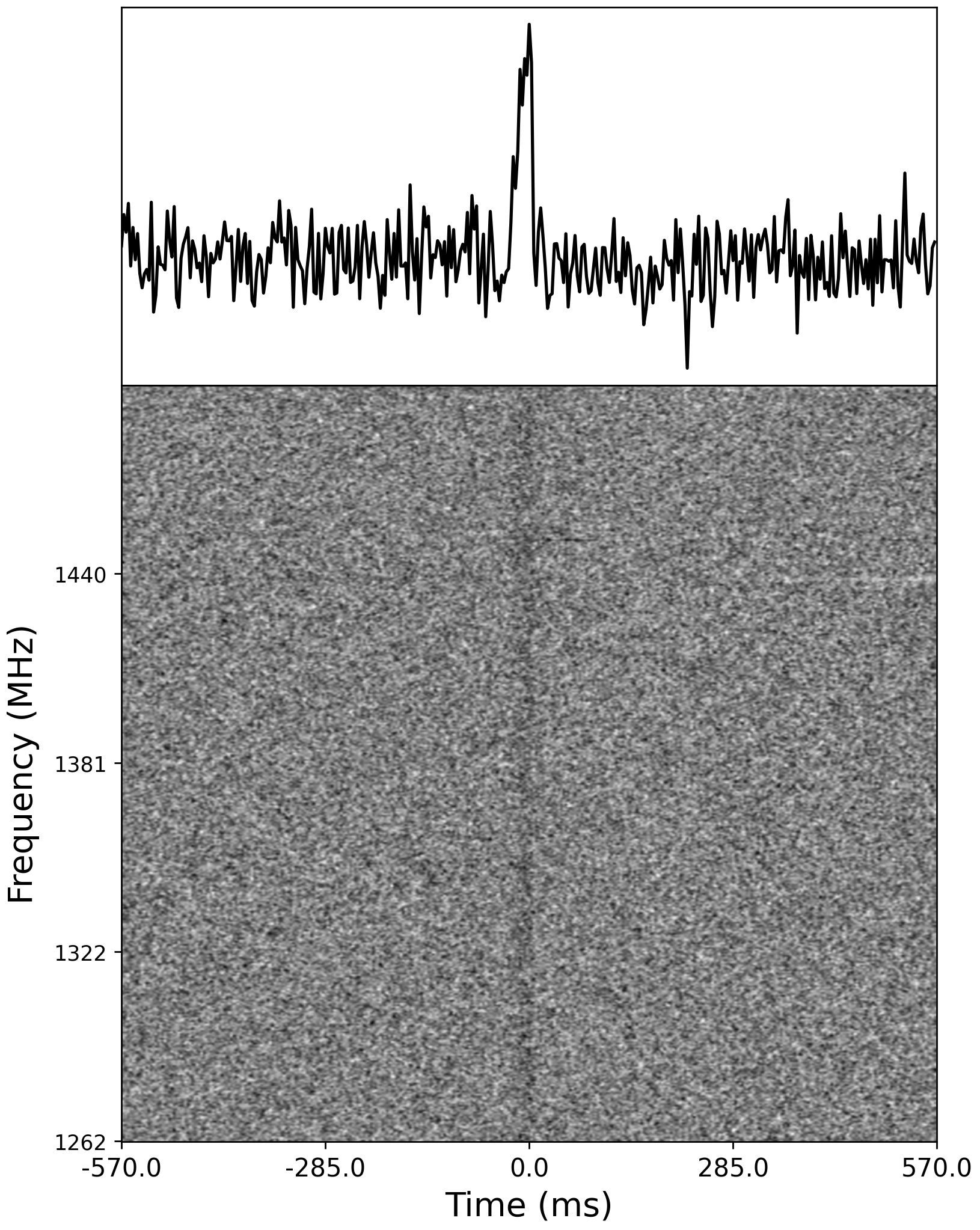} &
\includegraphics[width=5.2cm,angle=0]{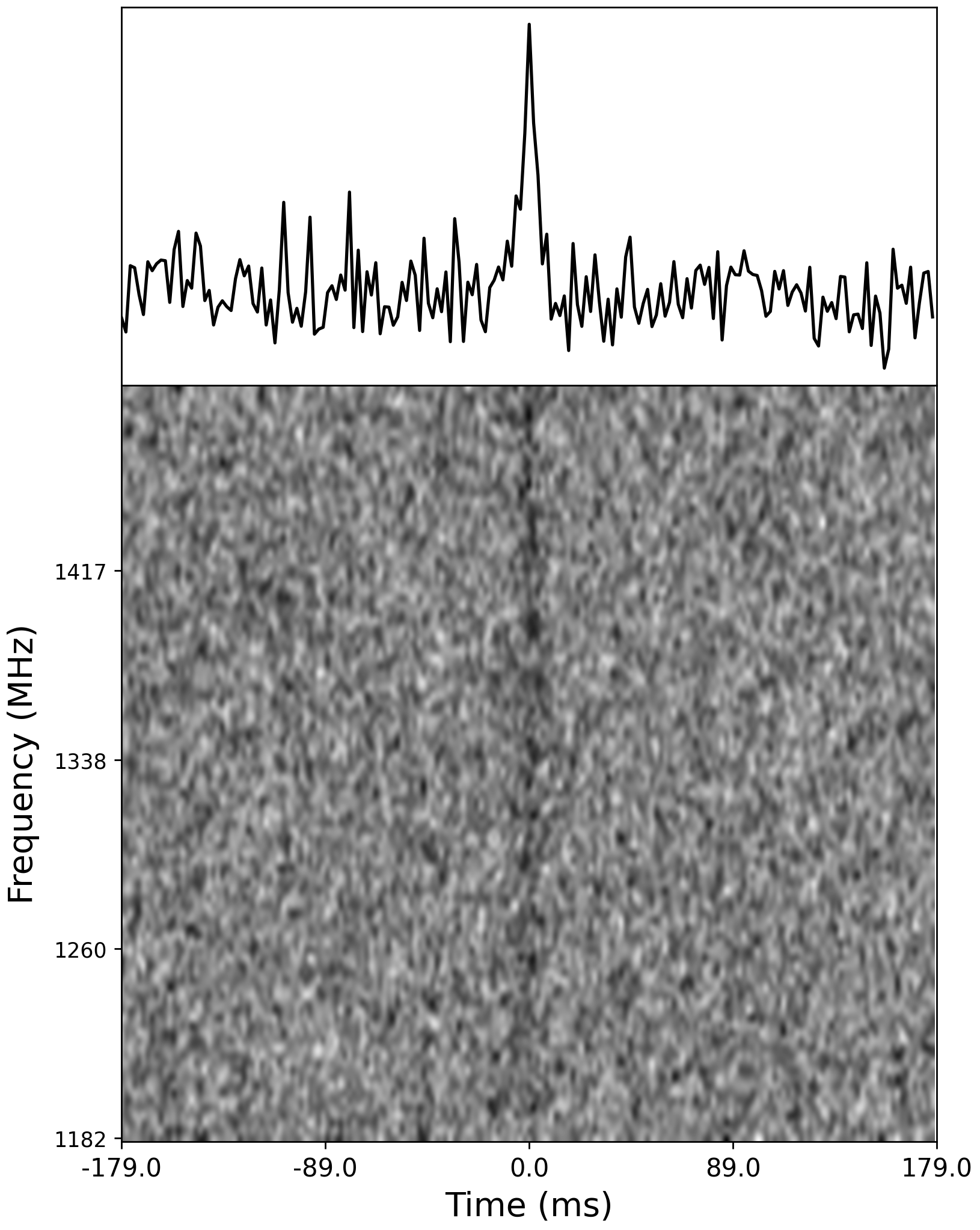} \\
(a) SPC001122 & (b) SPC991113 & (c) J1709$-$43 \\
\end{tabular}
\caption{Profiles and dynamic spectra for examples of detected pulses from three repeating sources. The bursts are plotted after being de-dispersed using DMs of 127.2 pc\,cm$^{-3}$ for (a) SPC001122, 186.4 pc\,cm$^{-3}$ for (b) SPC991113, and 225.8 pc\,cm$^{-3}$ for (c) J1709$-$43, respectively. (a) is the Parkes UWL data, while (b) and (c) are the Multibeam data.}
\label{figure:repeating}
\end{center}
\end{figure*}

After re-analysing the results of the 13 RRATs and 29 FRBs, we identified three RRATs, SPC 991113, J1709$-$43 and J0808$-$32, as likely repeating sources. Figure~\ref{figure:repeating} presents examples of our detected repeating pulses from each of these three sources.
Due to their similar DM and close positions, we suggest two of them, SPC 991113 and J1709$-$43, likely originate from the same sources as two other reported repeating sources, RRAT J1739$-$2521~\citep{Cui17} and PSR J1709$-$4401~\citep{Bates12}, respectively. 
Table~\ref{table:repeating} lists the properties of these previously reported one-off detection and their related repeating sources. The DMs of the related repeating sources are consistent with those of the one-off detections, and the differences in localization between the reported one-off detections and the related repeating sources are well within the localization uncertainties of the one-off detections.

From our 15.62\,hr UWL observation of J0808$-$32 at the pointing (J2000) 08:08:10.6, $-$32:18:11.0~\citep{Zhang20}, we detected a total of 34 repeating single pulses, leading to a pulse rate of $\sim$ 2.2\,hr$^{-1}$. 
%
By analysing the greatest common divisor of the time intervals between the time of arrival (TOA) of the detected pulses, we identified a period of 1.297047(5)\,s by minimizing the variance of the expected TOA phases.
However, no significant signal was observed from the periodic search pipeline using the \emph{\sc presto} package~\citep{Ransom01, Zhang18}, nor by manually folding each observation using the {\sc dspsr} package~\citep{Straten11} based on this derived period. 
The best estimate of the RM for the detected single pulses yielded a value of 77.2(2) rad m$^{-2}$. 
The Galactic longitude and latitude of J0808$-$32 are approximately 245.0$^\circ$ and 0.2 $^\circ$, respectively, and its DM of around 
127.2 pc\,cm$^{-3}$ indicates a relatively small distance of $\sim$ 433\,pc using the YMW16 model~\citep{Yao17}, and $\sim$ 1582\,pc using the NE2001 model~\citep{ne2001}.
More observations are encouraged to investigate this source further.

\subsection{Apparently one-off sources}\label{sec:results_non}

\begin{figure*}
\centering
\begin{tabular}{cc}
\includegraphics[height=6.5cm]{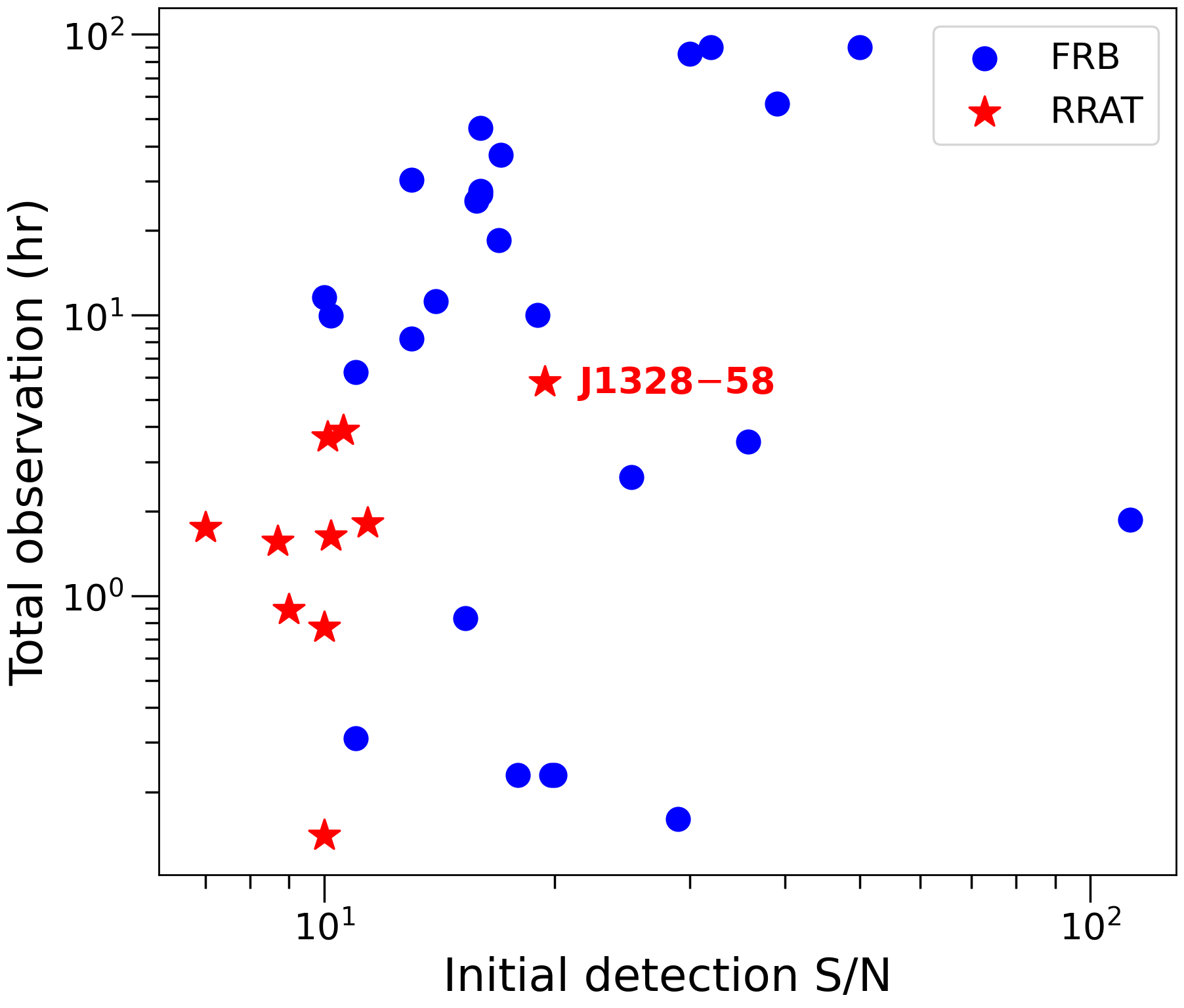} &
\includegraphics[height=6.5cm]{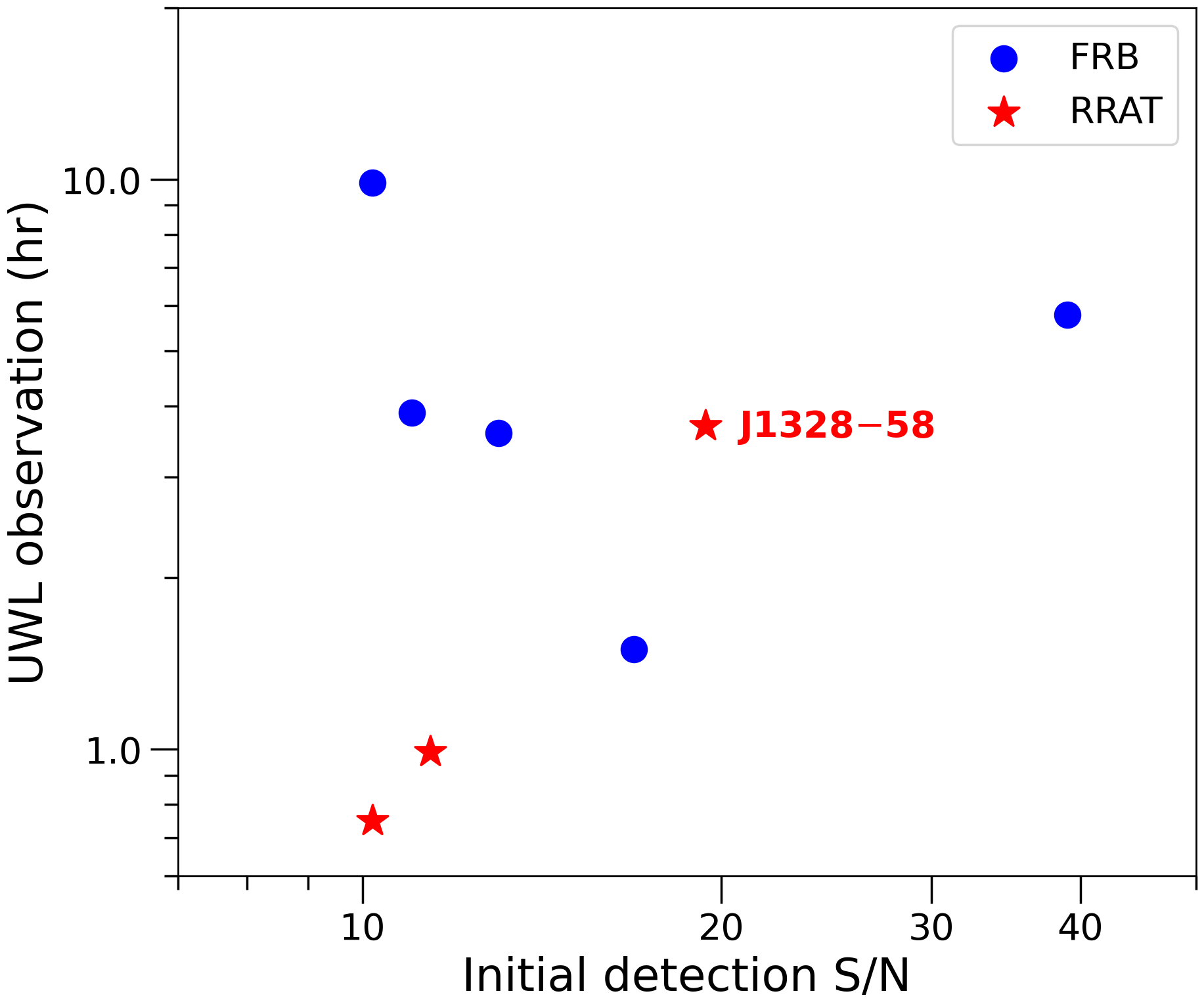} \\
\end{tabular}
\caption{The initial detection S/N versus the archival total observations (left) and UWL observations (right). The FRBs and RRATs are represented by the blue circle and red star, respectively.}
\label{fig:snr_obs}
\end{figure*}

The details of our accessible archival follow-up observations of the other 10 RRATs and 29 FRBs are listed in Table~\ref{tab:property}.
For three FRBs, FRB110220A, FRB110626A and FRB110703A, we did not find any archival datasets. As indicated in Table~\ref{tab:property}, we re-detected the initial pulses for 7 RRATs and 8 FRBs, and provided the precise pointing centres for the detected beams, S/N and DMs from our processing.   
However, no repeating signals were detected from these datasets. 

We presented the initial detection S/N versus the archival total observations for these apparently one-off RRATs and FRBs in the left panel of Figure~\ref{fig:snr_obs}.
In principle, since our single pulse search threshold is 7, a more significant
initial detection and 
further
follow-up observations would place more strict constraints on the source repeatability, and indicate a ``one-off'' nature.
From the total observations, the repeatability of the apparently one-off FRBs is better constrained, as most of them were detected with high S/N and follow-up observed with relatively sufficient time. 
One RRAT, J1328$-$58, with the highest S/N of 19.4, compared to other RRATs with S/N up to 11.4, and also having the longest follow-up time of 5.78\,hr among all RRATs, appears to be the most reliable one-off RRAT source.
%
However, it is notable that some sources have not been adequately re-observed to constrain the detectability,
with 3 RRATs and 6 FRBs observed for less than even 1\,hr.

Higher sensitivity follow-up observations are expected to detect fainter and more events from a repeating source. Based on the radiometer equation~\citep{Lorimer04handbook}, the sensitivity of UWL observation is about 1.7~\footnote{This considers the frequency range of 0.7-1.5\,GHz, which fully covers the FOV of the Multibeam observation.} to 3.4~\footnote{This considers the full frequency range of the UWL receiver.} times of that of the Multibeam observation. Assuming a power-law energy distribution like the repeating FRBs~\citep{Li21} and pulsar giant pulses~\citep{Bera19} with an index of $\sim -3$, the same length of UWL observation is expected to detect $\sim$ 5 to 39
times of the number of single pulses from the Multibeam observation. 
However, only 3 of the 10 RRATs and 5 of the 29 FRBs have been observed by the Parkes UWL, as shown in the right panel of Figure~\ref{fig:snr_obs}. 
Among these 8 sources, the UWL observing lengths for 3 are around 1\,hr, with the maximum UWL follow-up being less than 10\,hr for FRB010305A. Notably, RRAT J1328$-$58 still has the maximum UWL follow-up observations among these 3 RRATs.

\section{Discussion and Conclusions} \label{sec:con}

Among the 13 reported apparently one-off RRATs discovered by the Parkes telescopes, our re-analysis suggests three of them are repeaters: two (SPC 991113 and J1709$-$43) are previously known sources (J1739$-$2521 and J1709$-$4401), and one (J0808$-$32) is a new repeater.  
This repeating fraction is larger than that of the 30 FRBs detected by Parkes, where only one (FRB 180301A) has been observed to repeat by the FAST telescopes~\citep{Luo20}. Parkes UWL observation then also detected repeating pulses from this source~\citep{Kumar23}. 
As shown in Figure~\ref{fig:snr_obs}, even detected with larger S/N and followed by longer observations, FRBs are still harder to observe repeating signals. 
However, it is notable that many FRBs
discovered by the Parkes telescope
were still poorly followed, and only a few of them have been tracked with higher sensitivity observations such as Parkes UWL.  
It is challenging to use single-dish telescopes to follow the apparently one-off sources detected by single dishes: the FOV of a larger telescope normally cannot well cover the initial detection's localization, and a smaller telescope cannot provide sufficient sensitivity. 
Applying Parkes UWL to follow the apparently one-off sources detected by its Multibeam observations is efficient due to 
its higher sensitivity, and its FOV is larger than the localization uncertainties obtained by a single beam of the Multibeam receiver.
Our new detection of repeating pulses from J0808$-$32 and the detection of repeating pulses from FRB 180301A by~\cite{Kumar23} have proved the feasibility of this approach.
A better method would be to observe these sources using array telescopes such as MeerKAT~\citep[e.g., TRAPUM Survey,][]{Stappers16,Chen21},
which has even higher sensitivity, larger FOV, and can precisely localize the source immediately after a repeating signal is detected.

Besides searching for repeating signals from those apparently one-off sources, it is also important to investigate why they appear to be one-off in current observations. 
FRBs have a large sample of apparently one-off sources~\citep{CHIME21} and many models have been proposed to explain them~\citep{Platts19}.
However, the discussion of one-off Galactic radio pulses is lacking, and the relative sample size is potentially caused by the bias of the reported single pulse Galactic events~\citep{Lorimer04handbook, Keane16}.
We have presented 10 Galactic radio events that are still one-off, including one source detected with S/N of 19.4 that has been followed up with 5.78\,hr of observations, including 3.69\,hr of UWL time.

Several reasons could make a repeating Galactic source observed with only a single detection: (1) The initial observations have short durations, and the signals were affected by intense RFI or severe scintillation~\citep{Burke-Spolaor10}. (2) The source is extremely nulling and has a wide range of energy distribution, making only the very rare bright pulses detectable~\citep{Zhang23}. (3) The more frequent pulses are extremely faint and hard to detect, similar to the normal emission of pulsars with giant pulses~\citep{Popov06, Geyer21}, (4) some special RRATs (e.g., 1846$-$0257 and J1854$+$0306) exhibit extremely faint sequential emissions~\citep{Zhang24}, or (5) a magnetar like SGR~J1935+2154 can generate frequent pulses during specific periods~\citep{Zhu23}.
Scenario (1) could be well-examined by more similar observations. Observations with higher sensitivity are efficient to test scenarios (2) and (3), and necessary to test scenarios (4) and (5). 
Since the higher sensitivity is compared to the sensitivity of the initial detection, well-localized apparently one-off sources detected by a relatively low-sensitive array telescope, such as the ASKAP telescope~\citep{Bannister19, Wang24}, would be excellent to follow-up by other telescopes. 
The forthcoming Cryogenic phased array feed (CryoPAF) receiver of the Parkes telescope could also provide a good chance to detect numerous apparently one-off sources. 
After a large sample of one-off RRATs is obtained, and the details of non-detection observations of these sources are reported, catastrophic scenarios are worth proposing not only for extragalactic sources such as FRBs but also for Galactic sources. Such scenarios for close sources would be more practical to examine.

\section*{Acknowledgments}
We would like to express our gratitude to Lawrence Toomey for providing convenient and efficient downloading for the DAP datasets, which was invaluable for our research. This work is partially supported by the National SKA Program of China (2022SKA0130100), the National Natural Science Foundation of China (grant Nos. 12041306, 12273113,12233002,12003028), the international Partnership Program of Chinese Academy of Sciences for Grand Challenges (114332KYSB20210018), the National Key R\&D Program of China (2021YFA0718500), the ACAMAR Postdoctoral Fellow, China Postdoctoral Science Foundation (grant No. 2020M681758), and the Natural Science Foundation of Jiangsu Province (grant Nos. BK20210998). \\


\bibliography{sample631}{}
\bibliographystyle{aasjournal}



\end{document}